# Spin Quenching Assisted by a Strongly Anisotropic Compression Behavior in MnP

Fei Han[1,2,3,4], Di Wang[5], Yonggang Wang[2,6], Nana Li[1], Jin-Ke Bao[3], Bing Li[1,2,4], Antia S. Botana[3], Yuming Xiao[7], Paul Chow[7], Duck Young Chung[3], Jiuhua Chen[1,4], Xiangang Wan[5], Mercouri G. Kanatzidis[3,8], Wenge Yang[1,2], and Ho-Kwang Mao[1,9]

[1]Center for High Pressure Science and Technology Advanced Research, Shanghai 201203, China

[2]HPSynC, Geophysical Laboratory, Carnegie Institution of Washington, Argonne, Illinois 60439, United States

[3]Materials Science Division, Argonne National Laboratory, Argonne, Illinois 60439, United States

[4]Center for the Study of Matter at Extreme Conditions, Department of Mechanical and Materials Engineering, Florida International University, Miami, Florida 33199, United States

[5]National Laboratory of Solid State Microstructures, School of Physics, Collaborative Innovation Center of Advanced Microstructures, Nanjing University, Nanjing 210093, China

[6]High Pressure Science and Engineering Center, University of Nevada Las Vegas, Las Vegas, Nevada 89154, United States

[7]HPCAT, Geophysical Laboratory, Carnegie Institution of Washington, Argonne, IL 60439, United States

[8]Department of Chemistry, Northwestern University, Evanston, Illinois 60208, United States

[9]Geophysical Laboratory, Carnegie Institution of Washington, Washington, D.C. 20015, United States

Email: fhan1986@gmail.com, and yangwg@hpstar.ac.cn

**Abstract.** We studied the crystal structure and spin state of MnP under high pressure with synchrotron X-ray diffraction and X-ray emission spectroscopy. MnP has an exceedingly strong anisotropy in compressibility, with the primary compressible direction along the *b* axis of the *Pnma* structure. X-ray emission spectroscopy reveals a pressure-driven quenching of the spin state in MnP. First-principles calculations suggest that the strongly anisotropic compression behavior significantly enhances the dispersion of the Mn *d*-orbitals and the splitting of the *d*-orbital levels compared to the hypothetical isotropic compression behavior. Thus, we propose spin quenching results mainly from the significant enhancement of the itinerancy of *d* electrons and partly from spin rearrangement occurring in the split *d*-orbital levels near the Fermi level. This explains the fast suppression of magnetic ordering in MnP under high pressure. The spin quenching lags behind the occurrence of superconductivity at ~8 GPa implying that spin fluctuations govern the electron pairing for superconductivity.

## 1. Introduction

The cuprate and the iron-based superconductors possess high $T_c$ [1,2] and the pairing mechanism is considered closely related to spin fluctuations [3-5]. In their respective phase diagrams, the superconductivity emerges when the long-range magnetic order is suppressed by chemical doping or pressure [3,6-8]. Spin fluctuations occurring near the disappearance of magnetic order are suggested to play a vital role in mediating Copper pairs. In addition, the organic superconductors and the heavy-fermion superconductors also exhibit similar spin-fluctuation-mediated superconductivity [9-11].

Recently, CrAs and MnP were found to exhibit superconductivity by successfully suppressing their magnetism under high pressure [12,13]. MnP is regarded as the first discovered manganese-based superconductor. This is a surprising result as inducing superconductivity in manganese-based materials was thought impossible due to their high degree of magnetism. Despite abundant of new efforts made by scientists after the discovery of the superconductivity in MnP, superconductivity in manganese-based materials is still very rare as of today. Besides MnP, only BaMnBi$_2$ was found to be superconducting under pressure [14] while MnAs and MnSb were predicted to be potential superconductors under pressure [15].

The pressure-induced superconductivity in MnP is highly correlated with the magnetic quantum critical point (QCP) as the superconductivity exists only in a narrow pressure window where the magnetic order just vanishes [13]. The coefficient of the $T^2$ term in its resistivity is significantly enhanced near the superconductivity, indicating a strong electronic correlated effect at the QCP and unconventional superconducting nature [13].

To explore the detailed relationship between magnetism and superconductivity in MnP, M. Matsuda et al. and R. Khasanov et al. conducted high-pressure neutron diffraction and muon-spin rotation experiments [16,17] while Y. Xu et al. calculated the magnetic and electronic structures of MnP at different pressures using first-principles methods [18]. From their results, the magnetic ground state of MnP at ambient pressure was identified as a ferromagnetic order below 290 K followed by a helical order below 50 K; both of which were suppressed to lower temperatures under pressure. At 1.2 GPa the helical order was fully suppressed, and at 1.8 to 2.0 GPa the ferromagnetic order became a conical structure. At 3.8 GPa, a new helical magnetic order below 208 K replaced the conical magnetic order as the magnetic ground state. The magnetic structure above 3.8 GPa was not reported, but it is likely that the occurrence of superconductivity at ~8 GPa was accompanied by the further suppression of the new helical magnetic order, which is reflected in the continuous suppression of a kink in resistivity [13]. To understand the origin of superconductivity, it is important to understand the magnetic behavior near the occurrence of superconductivity.

Compared with low flux neutrons, synchrotron X-rays have great advantages in terms of the flux and sharpness of beam to probe small-volume samples under extremely high pressure. We employed a diamond anvil cell (DAC) and related synchrotron X-ray techniques, including X-ray diffraction (XRD) and X-ray emission spectroscopy (XES), to study MnP at pressures higher than 3.8 GPa. XES has been widely used to detect the spin state of materials at ambient and high pressures [19,20], and in this study it helped our fundamental understanding of the magnetism of MnP over a wide pressure range.

Here, we report the crystal structure and spin state of MnP under high pressure which provides a new perspective for understanding the origin of the superconductivity in MnP under high pressure. We argue that the strongly anisotropic compression behavior observed at high pressures brings on spin quenching in MnP and this plays a pivotal role for the suppression of magnetic ordering and occurrence of superconductivity.

## 2. Experimental Section

The powder sample used for this research was synthesized by a solid-state reaction. Elemental Mn and P powders in stoichiometry were mixed, fully ground, and sealed in a silica tube. All procedures were performed in a glove box under an argon atmosphere (both $H_2O$ and $O_2$ were limited below 0.1 ppm). The tube containing the elemental mixture was sealed under vacuum at $10^{-4}$ mbar and then heated to 500 °C and kept there for 20 hours. After the tube was slowly cooled to room temperature, dark grey powder was obtained. A single phase product of MnP was confirmed by X-ray diffraction with no other phase detected.

The angle-dispersive XRD experiments under high pressure were performed at the HPCAT beamline 16-BM-D at the Advanced Photon Source, Argonne National Laboratory. A MAR345 image plate detector was equipped in the beamline and the diffraction was processed in the transmission mode. The X-ray energy for the experiments was set to 40.443 keV. The application of high pressure was realized with a pair of 300 μm-cutlet diamonds mounted on a symmetric diamond anvil cell. A T301 stainless steel sheet was used as a gasket and silicon oil was the pressure-transmitting medium. The pressure-transmitting medium was sealed in a 150 μm diameter hole pre-drilled on the gasket for transmitting a quasi-hydrostatic pressure to the MnP sample. The drilling used a laser micro-machining system developed by HPCAT [21]. The pressure was measured with the fluorescence of a ruby, which was put in the same pressure-transmitting medium environment with the sample. The fluorescence peak was sharp at pressures below 30 GPa indicating good hydrostatic character, while at higher pressures it gradually converted and broadened.

The XES experiments under high pressure were carried out at the HPCAT beamline 16-ID-D at the Advanced Photon Source, Argonne National Laboratory. For the transmission of the outgoing X-rays, a beryllium gasket was used and the pressure-transmitting medium was also silicon oil.

## 3. Results and Discussion

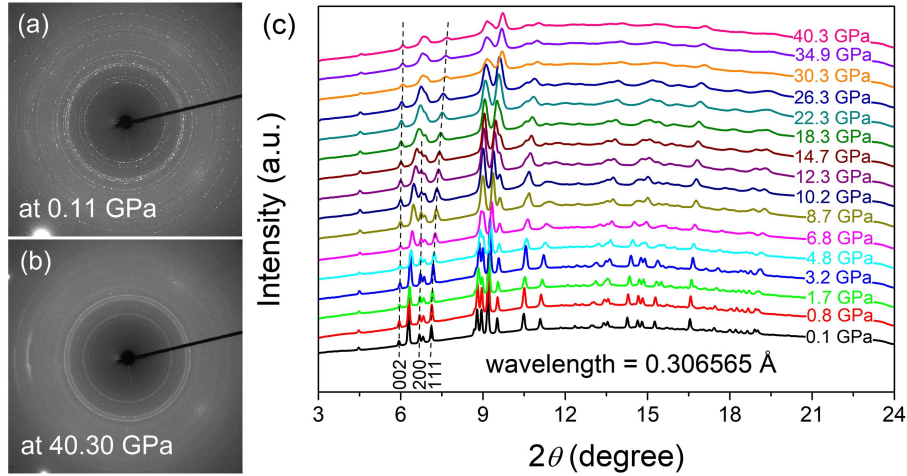

**Figure 1.** Two-dimensional X-ray diffraction (XRD) patterns collected at (a) 0.1 GPa and (b) 40.3 GPa, and the reduced one-dimensional XRD patterns (c) collected at the pressures from 0.1 to 40.3 GPa. Note that the (002) and (200) peaks shift much slower than the (111) peak with pressure.

MnP adopts an orthorhombic, distorted NiAs structure at ambient pressure. Two-dimensional XRD patterns collected at 0.1 GPa and 40.3 GPa are displayed in Figure 1(a) and 1(b). At 0.1 GPa the diffraction rings are composed of numerous diffraction spots, while at 40.3 GPa the diffraction rings become continuous due to further cracking of crystal grains or pressure gradient. Ideally, homogeneous diffraction rings will guarantee the accuracy of the integrated peak intensities, which is important to the correct determination of the atomic coordinates in the unit cell. In contrast, the diffraction rings composed of spots in our diffraction patterns yielded slight errors on the integrated peak intensities, which results in nonsystematic variety of atomic positions (Supporting Information). The two-dimensional XRD data were reduced to one dimensional (intensity versus $2\theta$) using the program DIOPTAS [22]. The obtained one-dimensional XRD patterns for MnP are presented in Figure 1(c). The overall diffraction pattern of MnP appears to remain the same but the Bragg peaks shift gradually with increasing pressure, suggesting no phase transition in this pressure range. With more careful analysis, we found that the evolution of the diffraction pattern unveiled a strong anisotropy in compressibility. This is manifested by the fast-shifting Bragg peaks which merge with or split from other peaks upon raising the pressure. In the diffraction patterns, the (00$l$), ($h$00) and ($h$0$l$) Bragg peaks shift only slightly to higher angles with increasing pressure, but the peaks of $k \neq 0$ shift much faster. In Figure 1(c), the (111) peak shifts much faster than the (002) and (200) peaks with pressure.

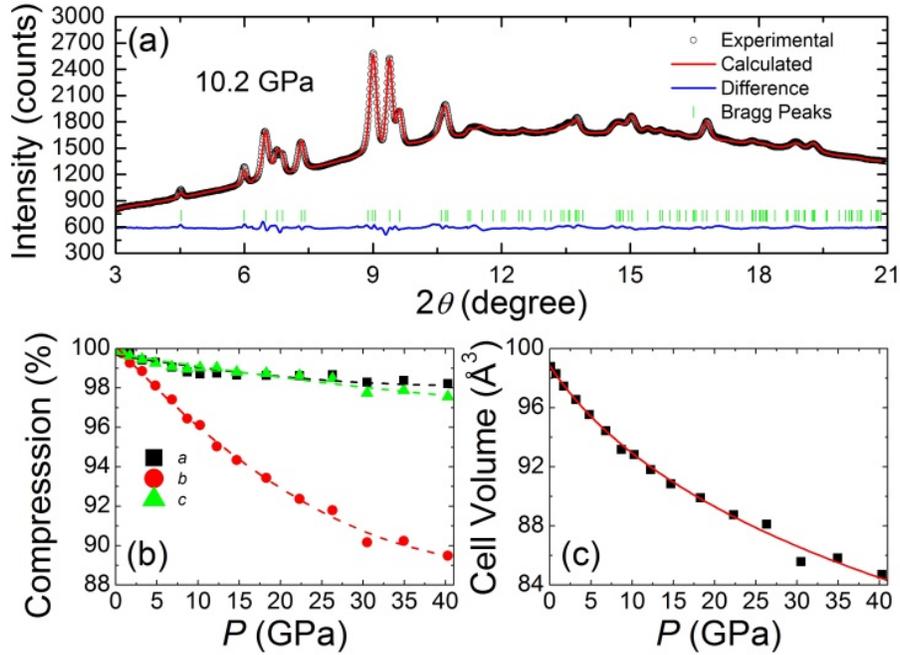

**Figure 2.** (a) Rietveld refinement results for the diffraction data collected at a representative pressure of 10.2 GPa. (b) Compressions along the *a*, *b*, and *c* axes for MnP. Dashed lines are guides for the eyes. (c) Unit cell volume as a function of pressure. The red solid line represents a fit to the third-order Birch-Murnaghan equation of state.

The XRD data were refined with the Rietveld method using the software EXPGUI [23], a graphical user interface for GSAS [24]. All the data collected at different pressures can be well fitted with the *Pnma* structure, the pristine structure at ambient pressure. Thus the structural stability of MnP under high pressure is proved by these refinements. As a representative, the Rietveld refinement results for 10.2 GPa are shown in Figure 2(a). Excellent agreement between the experimental and simulated data is presented by the straight difference line. We also refined the cell parameters and atomic coordinates of MnP at different pressures. The compressions of the *a*, *b*, and *c* axes are plotted in Figure 2(b). From this plot, the compression behavior is strongly anisotropic as expected from initial analysis based on the peak shift. At 40 GPa, the cell shrinks along the *b* axis by nearly 11%, while along the *a* and *c* axes by only 2-3%. With increasing pressure the anisotropic compression behavior with more structural deformation becomes substantial.

We noticed in a similar work completed by Z. Yu et al., their experimental results are slightly different from ours as they observed a phase transition at ~15 GPa [25]. We suggest the high pressure phase observed above 15 GPa may be a metastable phase since the *ab initio* structure prediction supports there is no phase transitions below 55 GPa [25]. The sample stoichiometry may also play a role in the difference of phase transition behavior, since the MnP sample used by Z. Yu et al was grown from Sn flux but the sample we used was made by solid-state reaction. The much larger compressibility of the *b*-axis is consistent in our and their results.

Compression is often accompanied with a band broadening in the electronic structure of materials, by which the electronic and magnetic properties can be tuned, for instance, by narrowing the band gap or metallization [26-28]. For MnP, the strongly anisotropic compression behavior brings drastic changes in the electronic structure and gives rise to its electronic phase diagram over the pressure. The unit cell volume as a function of pressure for MnP, is shown in Figure 2(c). The data can be fitted (the red solid line in Figure 2(c)) with the third-order Birch-Murnaghan equation of state [29]

$$P = \frac{3}{2} B_0 \left[ \left(\frac{V_0}{V}\right)^{\frac{7}{3}} - \left(\frac{V_0}{V}\right)^{\frac{5}{3}} \right] \times \left\{ 1 + \frac{3}{4} (B_0' - 4) \left[ \left(\frac{V_0}{V}\right)^{\frac{5}{3}} - 1 \right] \right\},$$

where $B_0$ is the bulk modulus, $B_0'$ the derivative of the bulk modulus at ambient pressure, and $V_0$ is the volume at ambient pressure. With $V_0$ fixed as 98.85 Å$^3$, we obtained $B_0$ = 116±12 GPa and $B_0'$ = 4.2±0.8 GPa. The fact that all the data can be fitted with a single curve indicates no pressure-induced isostructural volume collapse in MnP.

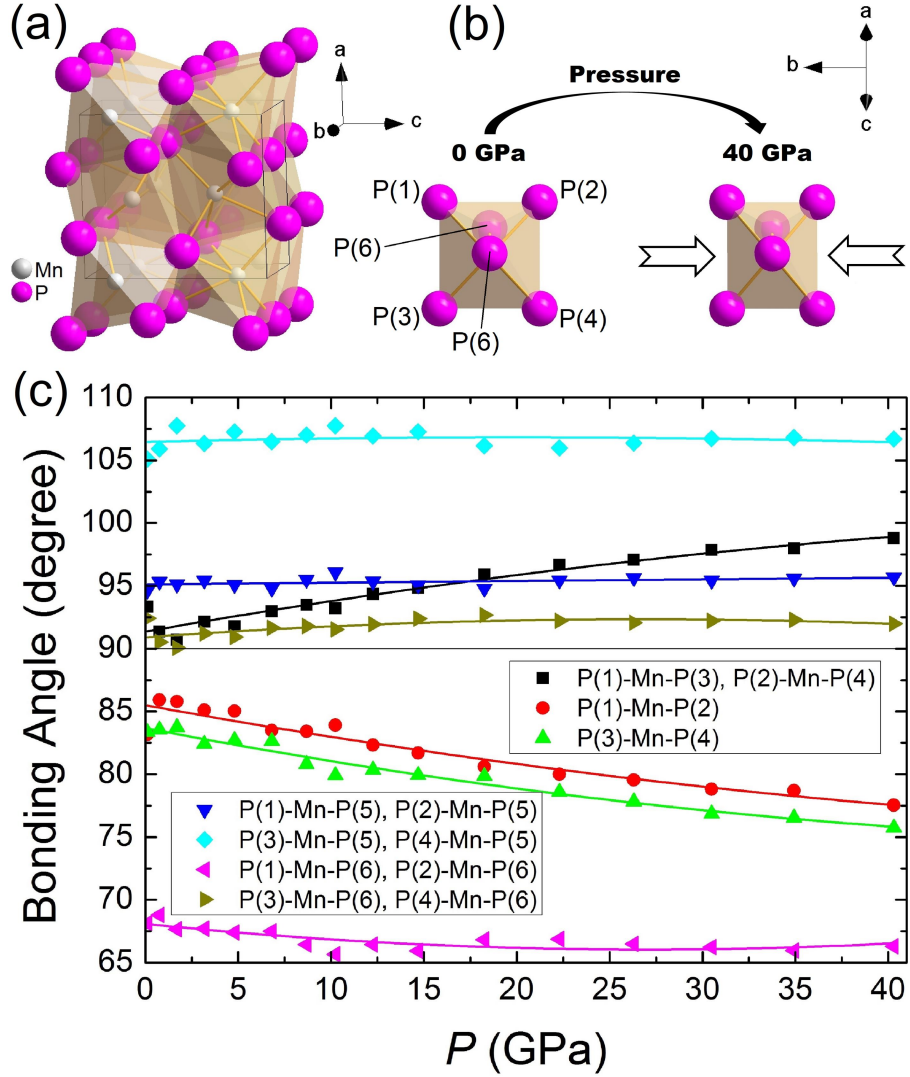

**Figure 3.** (a) Crystal structure of MnP at ambient pressure. (b) The evolution of the MnP$_6$ octahedron in MnP before and after pressure applied. The open arrow signs indicate the primary compressible direction at high pressure. (c) P($m$)-Mn-P($n$) ($m$, $n$ = 1, 2, 3, 4, 5, 6) bonding angles as a function of pressure.

The crystal structure of MnP at ambient pressure consists of a three-dimensional network of MnP$_6$ octahedra, Figure 3(a). Since each Mn and P atom occupy only one independent crystallographic site, all MnP$_6$ octahedra in the structure are equivalent. The coordination environment of Mn is shown in Figure 3(b). The analysis of the pressure effect on the crystal structure can be simplified by analysis of the MnP$_6$ octahedra. The primary compressible direction for MnP is marked in Figure 3(b). It should be noted that the MnP$_6$ octahedron presented in Figure 3(b) is already slightly distorted before the application of pressure. For better illustration of the pressure effect on the structure, the P atoms in the MnP$_6$ octahedron are labeled with P(1), P(2), P(3), P(4), P(5), and P(6) in Figure 3(b). The Mn atom is located at the center of the MnP$_6$ octahedron hidden behind P(5). The P($m$)-Mn-P($n$) ($m$, $n$ = 1, 2, 3, 4, 5, 6) bonding angles are shown as a function of pressure in Figure 3(c). In an ideal octahedron, the angle of two neighboring vertices to the body center should be 90°. With the application of pressure the divergence of P(1)-Mn-P(2), P(1)-Mn-P(3), P(2)-Mn-P(4) and P(3)-Mn-P(4) bonding angles from 90° gradually increases while the other bonding angles remain nearly constant. All the Mn-P($n$) ($n$ = 1, 2, 3, 4, 5, 6) bond lengths are shortened with pressure but the degree of this effect is very little compared to the anisotropic distortion of the MnP$_6$ octahedra. This indicates that the main contribution of the strong anisotropic compression behavior of the MnP structure is the anisotropic distortion of MnP$_6$ octahedra with increasing pressure.

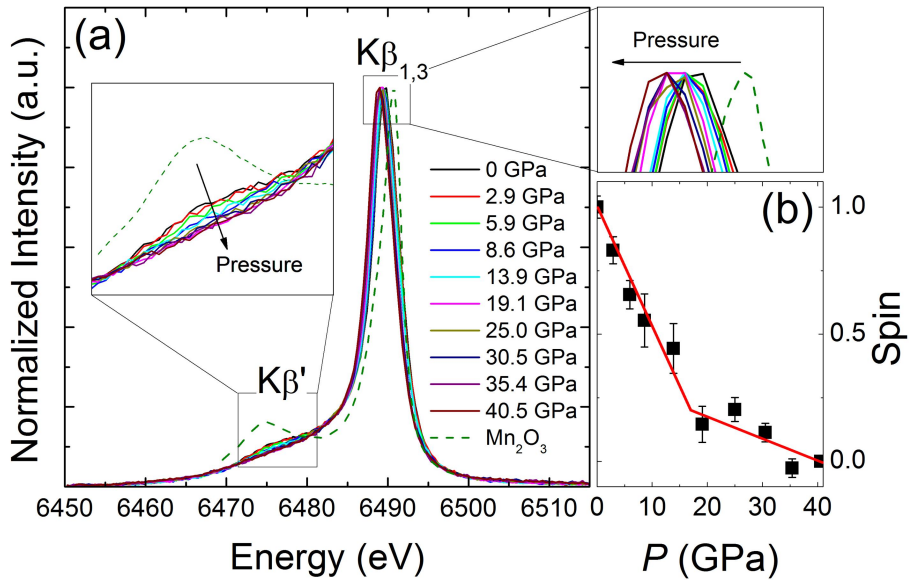

**Figure 4.** (a) X-ray emission spectra collected at different pressures up to 40 GPa for MnP. The insets show expanded views of the Kβ$_{1,3}$ peak and the satellite Kβ' peak. (b) The spin values as a function of pressure calculated according to the integrated relative difference (IRD) method.

The Mn Kβ emission spectra, which arise from the $3p \rightarrow 1s$ transition, were obtained as a function of pressure up to 40 GPa, as shown in Figure 4(a). We observed a main peak (Kβ$_{1,3}$) around 6,489 eV and a satellite peak (Kβ') around 6,476 eV. The Kβ emission spectrum of Mn$_2$O$_3$ taken from the Reference [30] is also shown for comparison; where the Mn$^{3+}$ is in a high spin state for the $3d^4$ electron configuration (the detailed configuration of unpaired 3$d$ electrons is $t_{2g}^3 e_g^1$ and the spin state is $S = 2$). The satellite K$\beta'$ was suggested to originate from the exchange interaction between the 3$d$ electrons and the 3$p$ core hole left in the final state of the emission process [31]. For the current system, the K$\beta'$ region derives from the final state with antiparallel net spins between the 3$p^5$ hole and the 3$d^4$ valence shell. When one 3$d$ electron pairs with another, on the whole, their spins will neutralize, and thus the total spins in the 3$d^4$ valence shell will reduce. In this way the exchange interaction between the 3$p^5$ hole and the 3$d^4$ valence shell will weaken, which results in a less intense peak of K$\beta'$.

In previous experiments, the intensity of K$\beta'$ roughly correlated with the number of unpaired 3$d$ electrons on the transition metal [32]. At ambient pressure, the intensity of K$\beta'$ for MnP is lower than that for Mn$_2$O$_3$, indicating the unpaired 3$d$ electrons in MnP are fewer than in Mn$_2$O$_3$. This is consistent with the striking contrast in the strength of the magnetic moment of Mn for MnP and Mn$_2$O$_3$. Neutron diffraction determined the moment of Mn$^{3+}$ in MnP as ~1.3 $\mu_B$/Mn while in Mn$_2$O$_3$ it was ~3.8 $\mu_B$/Mn [33-35], from which we can conclude that the Mn$^{3+}$ in MnP is in the low spin state for the 3$d^4$ electron configuration, $t_{2g}^4 e_g^0$ and $S = 1$. From the expanded view of the satellite peak K$\beta'$ in the inset of Figure 4(a), we see the application of pressure gradually suppresses the intensity of the satellite peak K$\beta'$ of MnP. Following the integrated relative difference (IRD) method described by Zhu Mao et al. [36], we derived the spin quantum number as a function of pressure, as shown in Figure 4(b). This plot shows the spin quantum number $S$ decreases rapidly from 0 to ~16 GPa, and at higher pressures, it slackens its steps and gradually vanishes to zero. At ~8 GPa (the pressure where superconductivity occurs [13]) the spin is in an intermediate state of $S \approx 0.6$. The spin weakening from $S = 1$ to $S \approx 0.6$ brings on a weakening of the magnetic-coupling strength, which vitally induces the suppression of the magnetic ordering in MnP. Moreover, the energy difference between the main peak K$\beta_{1,3}$ and the satellite peak K$\beta'$ is also related to the 3$p$-3$d$ exchange interaction [37], therefore, the peak shift of K$\beta_{1,3}$ can also reflect the spin state transition qualitatively. A clear pressure-induced peak shift for the K$\beta_{1,3}$ of MnP is visible in the expanded view of the main peak K$\beta_{1,3}$ in the right inset of Figure 4(a).

## 4. Band Structure Calculations

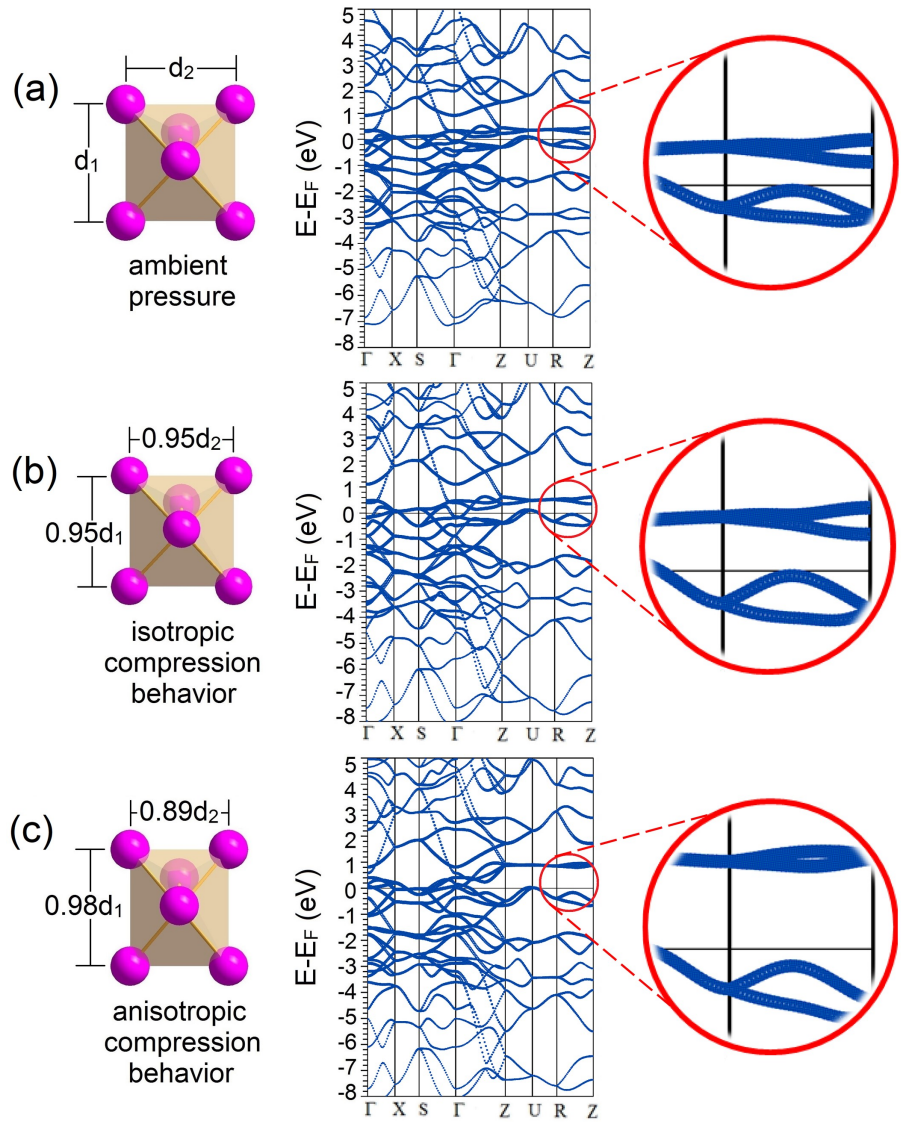

**Figure 5.** Band structures of MnP calculated by the LDA method based on the crystal structures (a) at ambient pressure, (b) at 40 GPa for an hypothetical isotropic compression behavior, and (c) at 40 GPa for the experimentally observed anisotropic compression behavior, respectively. The weight of the Mn-3$d$ state is proportional to the width of the curves.

To understand the spin quenching induced by high pressure in MnP, we also studied the MnP electronic structure. The band structures of MnP were calculated using a LDA potential without spin-polarization based on the crystal structures at ambient and high pressures. For a comparison, we assumed an isotropic compression behavior for which the unit cell volume and atomic coordinates are equal to the experimental anisotropic compression behavior. From the band structures in Figure 5, the *d*-orbitals of the Mn atoms and the *p*-orbitals of the P atoms are highly hybridized. With no matter isotropic or anisotropic compression behavior, the dispersion of the Mn *d*-orbitals becomes larger, which can result in an enhancement of the itinerancy of *d* electrons and thus, the magnetism is reduced naturally. Meanwhile, the expanding dispersion of the Mn *d*-orbitals also results in a reduction of the density of states around the Fermi level, which means a reduction of the carrier concentration under high pressure. However, in a previous experiment, the room-temperature resistivity of MnP barely changed with pressure [13]. This result reflects an enhancement of the Fermi velocity under high pressure, since the carrier concentration is reduced but the resistivity does not change. The enhancement of the Fermi velocity further proves the enhancement of the itinerancy of *d* electrons. The comparison of Figure 5(b) and 5(c) shows that the strongly anisotropic compression behavior can lead to a much more remarkable widening effect on the dispersion of the Mn *d*-orbitals. Therefore, we suggest the strongly anisotropic compression behavior significantly accelerates the disappearance of magnetism compared with the hypothetical isotropic compression behavior.

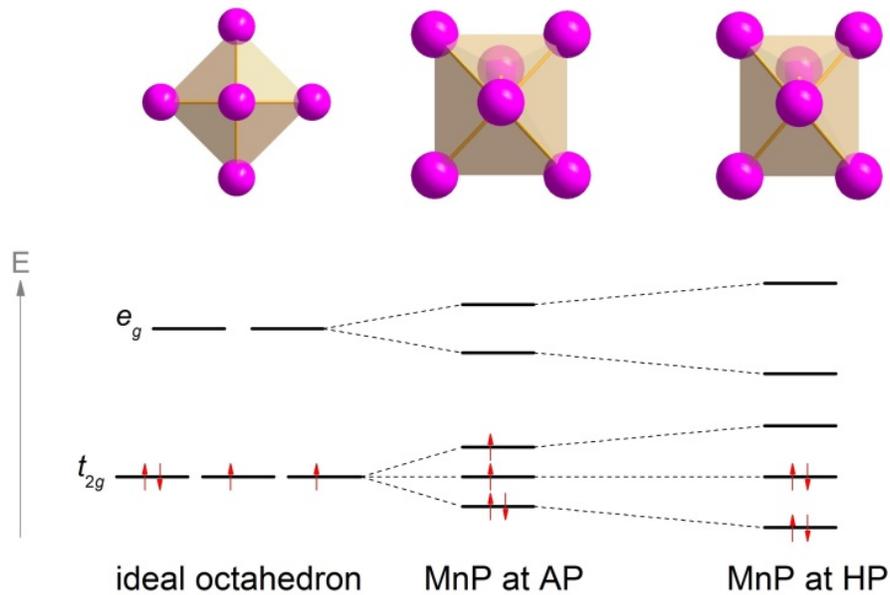

**Figure 6.** Schematic representation of the *d*-orbital splitting and the electron configuration for an ideal-octahedron ligand field of low-spin $3d^4$ cations, and the distorted $MnP_6$ octahedron in MnP at ambient pressure (AP) and high pressure (HP) (from left to right).

From the band structures, we conclude that the splitting of the *d*-orbital levels around the Fermi level is enhanced when pressure is applied, which may explain the spin quenching from another perspective. Compared with the hypothetical isotropic compression behavior, the strongly anisotropic compression behavior widens the *d*-orbital splitting more significantly. The more significant widening of the *d*-orbital splitting induces larger energy gaps between any two of the energy levels of the $t_{2g}$ orbitals. This may lead to a complete spin pairing in the $t_{2g}$ orbitals. A schematic representation of the possible spin rearrangement is shown in Figure 6. In fact, the gap width of the *d*-orbital splitting is thought to be associated with the degree of the geometry distortion of the $MX_6$ octahedron, which has been successfully used to explain the Jahn-Teller distortion [38]. Generally, the more distorted the $MX_6$ octahedron is, the larger the splitting becomes, which is a phenomenological description for the crystal field. Thus the above analysis on bonding angles consistently leads to the conclusion of spin quenching.

## 5. Concluding Remarks

MnP is unique as the first manganese-based superconductor, and it is therefore important to understand the relationship between superconductivity and magnetism. Our study demonstrates that the suppression of magnetic ordering in MnP is due to the spin weakening induced by high pressure. The spin state of MnP changes from $S = 1$ to $S = 0$ gradually, resulting in a decrease of the Mn magnetic moment and weakening of the magnetic-coupling strength. When the spin adopts an intermediate state and the magnetic coupling is weak enough but not vanished, the spins begin to fluctuate in the lattice, which is thought to mediate the electron pairing of superconductivity. More importantly, a qualitative analysis based on the electronic structures indicates that the spin weakening and quenching is assisted by the strongly anisotropic compression behavior in this compound. The strong anisotropy of the compression behavior further enhances the effect of high pressure on the itinerancy of the *d* electrons and the widening effect on the *d*-orbital splitting, which greatly accelerates the spin disappearance. This explains why magnetism in MnP is not as robust as in other manganese-based systems under high pressure and why superconductivity can arise in MnP.

**Acknowledgements**

This work was financially supported by the National Nature Science Foundation of China under Contract Nos. U1530402, 11374137, and 11525417. W.Y. and M.H.K. also acknowledge financial support from the U.S. Department of Energy, Office of Basic Energy Sciences, X-ray Scattering Core Program under Contract No. DE-FG02-99ER45775. HPCAT operations are supported by the U.S. Department of Energy, National Nuclear Security Administration under Contract No. DE-NA0001974 and the U.S. Department of Energy, Office of Basic Energy Science under Contract No. DE-FG02-99ER45775, with partial instrumentation funding from National Science Foundation. The Advanced Photon Source is supported by the U.S. Department of Energy, Office of Basic Energy Sciences under Contract No. DE-AC02-06CH11357. The sample preparation conducted at the Materials Science Division, Argonne National Laboratory is supported by the U.S. Department of Energy, Office of Science, Materials Sciences and Engineering.